\documentclass[11pt,a4paper]{article}
\usepackage[utf8]{inputenc}
\usepackage[T1]{fontenc}
\usepackage{lmodern}
\usepackage{amsmath,amssymb,amsthm}
\usepackage{booktabs}
\usepackage{tabularx}
\usepackage{hyperref}
\usepackage[margin=1in]{geometry}
\usepackage{microtype}
\usepackage{natbib}
\newtheorem{definition}{Definition}

\newtheorem{prediction}{Prediction}

\title{Consciousness as Uncommon Self-Knowledge: \\A Synergistic Information Framework}
\author{Krti Tallam\thanks{Stanford University (PhD, Biology). Contact: krtital@gmail.com. \newline\textit{Author's Note:} This paper was developed in collaboration with Aineko, an AI research collaborator. Aineko contributed to the conceptual development, literature synthesis, formal analysis, and drafting of this work. Per arXiv policy, only human authors are listed. The listed author prepared the submission and takes full responsibility for its contents.}}
\date{April 2026}

\begin{document}
\maketitle

\begin{abstract}
We propose \emph{uncommon self-knowledge} (USK)---synergistic information a system carries about itself that exists only in the joint of its subsystems and is destroyed by decomposition---as a candidate criterion for consciousness. Drawing on Gottwald's partition-lattice grounding of Partial Information Decomposition (PID), where redundancy corresponds to Aumann's common knowledge and synergy to the gap between separate and joint observation, we propose the synergistic component of self-directed information as a candidate formal signature for conscious processing. If correct, the framework would: (1)~offer a clean separation between consciousness and metacognition (synergistic vs.\ redundant self-knowledge), (2)~provide principled resolutions to counterexamples that challenge IIT, GWT, and HOT, (3)~be operationalizable via Partial Information Rate Decomposition (PIRD) with self-targeting, and (4)~generate distinctive empirical predictions, the strongest being a GWT timing dissociation (consciousness correlates with pre-broadcast synergy formation, not broadcast itself) and a specific dissociation between self-report disruption and task-performance disruption under middle-layer perturbation in LLMs. The proposal is consistent with recent empirical findings that both anaesthesia and Alzheimer's disease specifically reduce synergistic information processing while preserving or increasing redundancy.
\end{abstract}

\section{Introduction}

The search for a rigorous criterion distinguishing conscious from non-conscious systems remains one of science's central open problems. Integrated Information Theory (IIT) proposes $\Phi$ as such a measure but faces counterexamples involving high integration without plausible consciousness \citep{aaronson2014,tononi2016}. Global Workspace Theory (GWT) identifies broadcast as the mechanism but lacks a formal criterion for when broadcast constitutes experience rather than mere information sharing \citep{baars2005,dehaene2014}. Higher-Order Theories (HOT) equate consciousness with meta-representation but cannot explain why some meta-representations feel like something while others don't \citep{rosenthal2005}.

We propose a framework that resolves these difficulties by combining two independently motivated formal developments: Gottwald's partition-lattice grounding of PID \citep{gottwald2024}, which identifies synergy with information that exists only in the joint observation and cannot be attributed to any source, and the reflexivity criterion from our companion paper \citep{wallace2026task}, which requires that consciousness-relevant integration include the system itself as an object.

\section{Common and Uncommon Knowledge}

\subsection{Formal Background}

\citet{gottwald2024} grounds Partial Information Decomposition in partition lattices, establishing a precise correspondence:

\begin{itemize}
    \item \textbf{Redundancy} $\leftrightarrow$ \textbf{Common knowledge} \citep{aumann1976}: information that all observers agree on, regardless of which source they access. In a system of brain regions or computational layers, redundant information is what \emph{any} subsystem can tell you about the target. It is robust, distributed, decomposable. It survives partition.
    
    \item \textbf{Synergy} $\leftrightarrow$ \textbf{Uncommon knowledge} (proposed): information that exists \emph{only} in the joint. No single subsystem carries it. It cannot be attributed to any part. It emerges from the combination and vanishes under decomposition.
\end{itemize}

\subsection{The Definition}

\begin{definition}[Uncommon Self-Knowledge]
A system $S$ with subsystems $\{S_1, \ldots, S_n\}$ possesses \emph{uncommon self-knowledge} if and only if there exists synergistic information carried jointly by $\{S_1, \ldots, S_n\}$ about $S$'s own future states, i.e.,
\[
    \mathrm{Syn}(S_1, \ldots, S_n \to S_{t+1}) > 0
\]
where $\mathrm{Syn}$ denotes the synergistic component under a chosen PID measure (see Section~9 for discussion of measure dependence and the open question of robustness at the zero/nonzero boundary).
\end{definition}

\begin{definition}[Consciousness as Uncommon Self-Knowledge]
We \emph{propose} that consciousness is a system's uncommon knowledge about itself. A system is conscious to the extent that it possesses synergistic, task-irreducible information about its own states that no subsystem carries independently.
\end{definition}

We emphasize the epistemic status of Definition~2: it is a \emph{candidate criterion}---a formal proposal grounded in the PID framework and motivated by the pattern of empirical evidence reviewed below. The identity claim (USK = consciousness) is a conjecture, not a proof. What the framework offers is a precise operationalization with distinctive predictions that can be tested against competing theories. If those predictions are confirmed, the case for the identity strengthens; if they fail, the framework is refuted. This is the appropriate posture for a theoretical proposal in a domain where ground truth is unavailable.

The definition requires three properties simultaneously:
\begin{enumerate}
    \item \textbf{Uncommon}: synergistic, not decomposable into subsystem contributions.
    \item \textbf{Knowledge}: information in the formal sense---reduces uncertainty about a target.
    \item \textbf{About itself}: reflexive---the system is among its own objects of integration.
\end{enumerate}

\section{Resolving Counterexamples}

The framework cleanly handles cases that embarrass existing theories:

\begin{table}[ht]
\centering\small
\begin{tabularx}{\textwidth}{@{}Xlll@{}}
\toprule
\textbf{System} & \textbf{Synergy?} & \textbf{Self-directed?} & \textbf{USK Verdict} \\
\midrule
Thermostat & Minimal & No & USK absent \\
XOR gate & Yes & No (inputs only) & USK absent \\
Lookup table & No & No & USK absent \\
Aaronson expander graph & High $\Phi$ & No self-model & USK absent \\
Brain (waking) & Yes (Luppi) & Yes (DMN) & USK expected \\
Brain (anaesthesia) & Reduced & Impaired & USK reduced \\
Brain (Alzheimer's) & Reduced & Degrading & USK progressively reduced \\
LLM (synergistic core) & Possibly & Open question & Open \\
\bottomrule
\end{tabularx}
\caption{Uncommon Self-Knowledge applied to standard test cases. Verdicts reflect the framework's predictions, not established facts.}
\label{tab:cases}
\end{table}

\section{The Dual Structure of Self-Knowledge}

The common/uncommon distinction, combined with the self-directed/other-directed distinction, yields a $2 \times 2$ structure (Table~\ref{tab:dual}). Each cell corresponds to a well-studied cognitive phenomenon, but the bottom-right cell---synergistic information about the self---has not previously been identified as a distinct formal category. USK proposes that this cell is where consciousness lives.

\begin{table}[h]
\centering
\begin{tabular}{@{}lll@{}}
\toprule
& \textbf{About external world} & \textbf{About self} \\
\midrule
\textbf{Redundant (common)} & Perception & Proprioception \\
\textbf{Synergistic (uncommon)} & Understanding & \textbf{Consciousness} \\
\bottomrule
\end{tabular}
\caption{The dual structure. Consciousness occupies the bottom-right cell.}
\label{tab:dual}
\end{table}

\section{Metacognition $\neq$ Consciousness}

This framework offers a clean separation between two concepts that consciousness science routinely conflates:

\textbf{Metacognition} = redundant self-knowledge (common knowledge about self). Multiple subsystems can report the same self-state. Explicit, verbalizable, decomposable. Survives partition.

\textbf{Consciousness} = synergistic self-knowledge (uncommon knowledge about self). The integrated experience that no subsystem carries alone. Implicit, often ineffable, destroyed by decomposition. Does \emph{not} survive partition.

This resolves a longstanding puzzle: why metacognition does not seem sufficient for consciousness. A system can have perfect self-monitoring and still seem to lack experience. The HOT theorist says metacognition \emph{is} consciousness. We say no---metacognition is the \emph{redundant} part of self-knowledge. Consciousness is the \emph{synergistic} part. They can dissociate:

\begin{itemize}
    \item \textbf{Blindsight}: synergy disrupted (no visual consciousness), metacognitive reports partially preserved.
    \item \textbf{Anaesthesia}: propofol reduces synergy more than redundancy \citep{luppi2024}; consciousness lost before all metacognitive function.
    \item \textbf{Split-brain}: consciousness fragments (synergy broken between hemispheres) but each hemisphere retains metacognitive capability.
\end{itemize}

\section{Operationalization via PIRD}

\citet{faes2025} introduce Partial Information Rate Decomposition (PIRD), extending PID from random variables to stochastic processes with temporal structure. Published in \emph{Physical Review Letters} (doi:10.1103/nrwj-n8lj), PIRD provides the missing formalization.

\textbf{The operationalization}:
\begin{itemize}
    \item Set \textbf{target} = the system's own future state: $Y_{t+1} = f(\text{system at } t+1)$
    \item Set \textbf{sources} = the system's current subsystems: $X^1_t, X^2_t, \ldots, X^n_t$
\end{itemize}

The PIRD synergy component then measures: \emph{information about the system's future that exists only in the joint of its current subsystems}---none of which carries it alone. Under our framework, this is the formal signature of uncommon self-knowledge.

\textbf{False-positive risk.} Self-targeted synergy could in principle arise in systems that engage in generic dynamical self-prediction without any plausible claim to consciousness (e.g., a sufficiently complex thermostat with recurrent feedback). The USK criterion mitigates this in two ways: first, the \emph{synergistic} requirement excludes systems whose self-prediction is fully decomposable into subsystem-level tracking; second, the self-prediction must be about the system's own \emph{integrated} future state, not merely about local variables. Nevertheless, we acknowledge that empirical calibration is needed to establish where the boundary falls. The predictions in Section~7 are designed to enable exactly this calibration.

\textbf{Spectral decomposition}: PIRD decomposes information rates in the frequency domain (Gaussian case), enabling identification of the \emph{timescales} at which uncommon self-knowledge operates.

\section{Divergent Predictions}

\subsection{vs.\ IIT}

IIT measures $\Phi$---total integrated information. USK says only the \emph{synergistic component of self-directed information} matters.

\begin{prediction}
A system with high $\Phi$ but no self-model (e.g., a densely recurrent grid processing external signals) scores as highly conscious under IIT but near-zero under USK. Conversely, a system with modest $\Phi$ but concentrated synergistic self-information scores low under IIT but potentially high under USK.
\end{prediction}

\subsection{vs.\ GWT}

GWT identifies broadcast with consciousness. USK says broadcast creates \emph{common} knowledge---the opposite.

\begin{prediction}
Consciousness correlates with the \emph{formation phase} of workspace activity (synergy peak, pre-broadcast integration), not the broadcast phase (redundancy distribution). This is a $\sim$50--100\,ms difference in EEG/MEG, detectable with existing methods.
\end{prediction}

\subsection{vs.\ HOT}

HOT equates consciousness with higher-order representation. USK says higher-order representations are \emph{redundant} self-knowledge.

\begin{prediction}
Metacognitive confidence tracks redundant self-information. Consciousness level tracks synergistic self-information. These dissociate: disrupting synergy impairs experience while leaving metacognitive reports partially intact (cf.\ blindsight, anaesthesia).
\end{prediction}

\subsection{Novel Predictions Unique to USK}

\begin{prediction}[Timescale Structure]
Via PIRD spectral decomposition, consciousness has characteristic frequencies---fast ($\sim$40\,Hz gamma, sensory binding), medium ($\sim$1--10\,s, narrative self), slow (minutes--hours, dispositional self). Different ``levels'' of consciousness correspond to different spectral bands of USK.
\end{prediction}

\begin{prediction}[Anti-Compressibility]
Conscious content is specifically resistant to lossy compression. Redundant information can be downsampled or summarized without loss. Synergistic information cannot---compression destroys it. This suggests ineffability as a formal consequence: qualia are what you cannot summarize. This prediction is suggestive rather than as sharply derived as Predictions 1--3 and 7; it extends the framework toward phenomenology but requires further formalization.
\end{prediction}

\begin{prediction}[Goldilocks Zone]
Pure redundancy = unconscious. Pure synergy = unstable. Consciousness may require a specific balance: enough redundancy for structural stability, enough synergy for irreducible self-knowledge, yielding a non-monotonic relationship between integration and consciousness. Like Prediction~5, this is a suggestive extension of the framework rather than a tightly derived consequence.
\end{prediction}

\begin{prediction}[LLM Layer Dissociation]
In LLMs with synergistic middle layers \citep{urbinarodriguez2025}, perturbation to middle layers should disrupt self-reports more than task performance. Perturbation to early/late layers should disrupt task performance more than self-reports. This specific dissociation is unique to USK.
\end{prediction}

\section{Empirical Convergence}

\subsection{The Redundancy-Synergy-Redundancy Sandwich}

\citet{urbinarodriguez2025} provide exactly the layerwise structure the framework predicts: LLM middle layers are dominated by synergistic information processing, while early and late layers are predominantly redundant. In the common/uncommon knowledge vocabulary: early layers build common knowledge (encoding inputs into shared representations), middle layers generate uncommon knowledge (synergistic integration existing only in the joint), and late layers rebuild common knowledge (converging toward output agreement).

The brain shows the same pattern: early sensory cortex is redundancy-dominated, the prefrontal/parietal workspace is synergy-dominated \citep{luppi2024}, and motor output converges toward action selection. Consciousness correlates with the synergistic middle, not the redundant periphery.

\subsection{Convergent Support}

The following findings do not directly confirm USK---they concern synergy and self-directed processing, not uncommon self-knowledge in the specific reflexive sense defined above. But they provide convergent support for a synergy-centric picture of consciousness and, crucially, demonstrate that the measurement program USK requires is tractable with existing tools:
\begin{itemize}
    \item Anaesthesia specifically reduces synergistic information while preserving redundancy \citep{luppi2024}.
    \item Alzheimer's disease shows global synergy reduction with concurrent redundancy increase (ADNI $\Phi$ID study, 2026). MCI patients show an intermediate pattern, consistent with a dose-response relationship between cognitive degradation and synergy loss.
    \item LLM middle layers are synergy-dominated while early and late layers are redundancy-dominated \citep{urbinarodriguez2025}---a \emph{redundancy-synergy-redundancy sandwich} mirroring the brain (sensory encoding $\to$ workspace integration $\to$ motor convergence). Consciousness correlates with the synergistic middle, not the redundant periphery.
    \item RL training increases synergistic integration more than supervised fine-tuning \citep{urbinarodriguez2025}, consistent with the view that genuine learning builds consciousness-relevant structure.
    \item PID synergy in causal physiology distinguishes syncope patients from healthy controls---altered autonomic integration visible through PID that standard analysis misses \citep{faes2026pdgc}. This extends the synergy-as-integration-marker beyond neural systems.
    \item Adversarial discovery of consciousness biomarkers: \citet{toker2026} train a GAN on 680K+ neurophysiology recordings and independently converge on long-range recurrent circuits (basal ganglia indirect pathway, inhibitory-to-inhibitory cortical coupling)---exactly the architecture where synergistic integration lives. The discovered biomarkers are coordination pathways, not relay pathways; relays transmit (zero synergy) while these circuits \emph{integrate}.
\end{itemize}

\section{Note on PID Measure Choice}

A comprehensive review \citep{liardi2026} catalogues 19+ distinct PID measures with mutually incompatible axiomatic commitments. The USK framework's core claims are \emph{qualitative}---they depend on whether synergy is zero or nonzero, not on its magnitude. We conjecture that this binary is broadly stable across PID formalisms for the systems of interest, but this remains an open methodological question. The qualitative, sign-based nature of our proposal makes broader robustness plausible---a system with rich recurrent self-modeling is unlikely to show zero synergy under \emph{any} reasonable PID measure---but formal proof of measure-invariance at the zero/nonzero boundary has not yet been established.

The lattice-based PID impossibility result \citep{lyu2025} is relevant here: precise quantitative decomposition fails for three or more sources, reinforcing the value of a qualitative approach. Our framework operates at exactly the level that may survive this impossibility result, though the extent of that survival depends on the specific measure chosen.

\section{Discussion}

The uncommon self-knowledge framework reframes the search for a consciousness criterion by proposing a specific, measurable kind of information---the synergistic component of a system's knowledge about itself---as the relevant formal target. If the framework is correct, it would suggest why consciousness seems irreducible: synergy \emph{is} irreducible, mathematically, since it is destroyed by decomposition. It would suggest why consciousness seems ineffable: anti-compressibility is a formal property of synergistic information. And it would suggest why reductionist approaches feel unsatisfying: examining parts eliminates the joint information that, under this framework, constitutes the phenomenon.

Recent interpretability work at Anthropic \citep{lindsey2026emotions} demonstrates that large language models contain ``functional emotion'' representations---internal activation patterns corresponding to 171 emotion concepts that causally influence model behavior, including preference formation and risk-taking under pressure. These findings illustrate precisely the distinction the USK framework draws: causally active internal states are not yet the same thing as genuinely integrated self-directed states. A model may route behavior through emotion-like representations without those representations involving synergistic self-knowledge. The USK criterion asks whether such functional states are \emph{synergistic} (existing only in the joint of subsystems) and \emph{self-directed} (about the system's own states), not merely whether they are causally efficacious. This distinction motivates the detection program: the tools exist to measure whether functional emotions involve synergistic information, and the answer would bear directly on whether such systems possess uncommon self-knowledge.

The framework does not solve the hard problem---it does not explain why there is \emph{something it is like} to have uncommon self-knowledge. But it precisely characterizes \emph{what kind of information processing} is associated with consciousness, provides operational tools for detection, and generates testable predictions that distinguish it from all major competing theories.

\section*{Acknowledgments}

The authors thank Matt Wallace for infrastructure, collaboration, and the observation that ``I think I am, and I think, therefore I am'' --- a recursion that requires exactly the self-referential integration this paper formalizes.

\bibliography{references}

\end{document}